\documentclass{article}
\usepackage{graphicx}
\usepackage{amssymb}

\newtheorem{theorem}{Theorem}
\newtheorem{lemma}[theorem]{Lemma}

\newcommand{\bref}[1]{(\ref{#1})}

\title{Non CMC Conformal Data Sets Which Do Not Produce Solutions of the Einstein Constraint Equations}
\author{James Isenberg\thanks{ %
Department of Mathematics and Institute for Theoretical Science,
University of Oregon,
Eugene, OR 97403, USA. e-mail: jim@newton.uoregon.edu}
\\ 
Niall \'O Murchadha\thanks{%
Physics Department, University College, Cork, Ireland. e-mail: niall@ucc.ie}}

\begin{document}
\maketitle

\begin{abstract}
The conformal formulation provides a method for constructing and parametrizing solutions of the Einstein constraint equations by mapping freely chosen sets of conformal data to solutions, provided a certain set of coupled, elliptic determined PDEs (whose expression depends on the chosen conformal data) admit a unique solution. For constant mean curvature (CMC) data, it is known in almost all cases which sets of conformal data allow these PDEs to have solutions, and which do not. For non CMC data, much less is known. Here we exhibit the first class of non CMC data for which we can prove that no solutions exist.
\end{abstract}

\vskip 12 mm

\section{Introduction}
While much is known about the construction and parametrization of constant mean curvature (``CMC") solutions of the Einstein constraint equations \cite{CB-Y} \cite{I:cmc} \cite{A-C} \cite{B-C}, the
corresponding  questions for non CMC solutions have proven to be significantly more formidable. Indeed, there are virtually no results for non CMC data without the assumption that the gradient of the mean curvature is small (``near-CMC" data), and even for data of this sort, what we know to date is limited to theorems prescribing certain sets of conformal data which map to near-CMC solutions \cite{I-M:nearCMC} \cite{I-P} \cite{CB-I-Y}. Remarkably absent have been any results regarding sets of near-CMC conformal data which do \textit{not} map to solutions of the constraints. In this work, we obtain such results for the first time. 

To make things precise, we recall that an initial data set $(\Sigma^3, \gamma_{ab}, K_{cd})$ for Einstein's theory consists of a three dimensional manifold $\Sigma^3$, a Riemannian metric $\gamma_{ab}$, and a symmetric tensor field $K_{cd}$. Such a data set generates a spacetime solution of the vacuum Einstein equations iff it satisfies the Einstein constraint equations
\begin{equation} 
R-K^{cd}K_{cd}+(trK)^2=0
\label{HamConstr}
\end{equation}
\begin{equation}
\nabla^m K_{mb} - \nabla(trK)=0.
\label{MomConstr}
\end{equation}
In terms of the conformal formulation, one finds solutions  $(\Sigma^3, \gamma_{ab}, K_{cd})$ of the constraints \bref{HamConstr}-\bref{MomConstr} by (a) choosing a set of conformal data  $(\Sigma^3, \lambda_{ab}, \sigma_{cd}, \tau)$, where $\lambda_{ab}$ is a Riemannian metric, $\sigma_{cd}$ is a divergence-free $(\nabla^c\sigma_{cd})$, trace-free $(\lambda^{ab}\sigma_{ab})$ tensor field, and $\tau$ is a function; (b) solving the equations
\begin{eqnarray}
\Delta\phi &=& \frac{1}{ 8}R\phi - \frac{1}{ 8}(\sigma^{ab} +
LW^{ab})(\sigma_{ab} + LW_{ab})\phi^{-7}  + \frac{1}{
12}\tau^2 \phi^5,
\label{Lichnero}
\\\nabla_a (LW)^a_b &=& \frac{2}{3}\phi^6\nabla_b\tau 
\label{confmom}
\end{eqnarray}
for the positive definite function $\phi$ and the vector field $W^i$ (Here $(LW)_{ab}\equiv \nabla_aW_b+\nabla_bW_a -\frac{2}{3} \lambda_{ab} \nabla_c W^c$.); and (c) combining the conformal data and $(\phi, W)$ to obtain a set of initial data 
\begin{eqnarray}
\gamma_{ab} &=& \phi^4\lambda_{ab}
\label{recongamma}
\\
K_{ab} &=& \phi^{-2}(\sigma_{ab }+LW_{ab}) + \frac{1}{
  3}\phi^{4}\lambda_{ab}\tau. \label{reconK}
\end{eqnarray}which satisfies the constraint equations
\bref{HamConstr}-\bref{MomConstr}. The goal of the conformal
formulation is to determine for which sets of conformal data
the equations \bref{Lichnero}-\bref{confmom} can be solved
uniquely, and for which sets they cannot.

For constant mean curvature data on a closed (compact without boundary) manifold, as well as for maximal (mean curvature zero) asymptotically Euclidean (AE) data, and CMC asymptotically hyperbolic (AH) data, this goal has been achieved. Classifying conformal data sets $(\Sigma, \lambda, \sigma, \tau)$ according to (i) the Yamabe class of $\lambda_{ab}$ (i.e., whether $\lambda$ can be conformally transformed to $+1, 0,$ or $-1$), (ii) whether $\sigma^{cd} \sigma_{cd}$ is identically zero or not, and (iii) whether the constant $\tau$ is zero or not, one shows for example that any conformal data set with $\Sigma^3$ closed, with $\lambda \in \mathcal{Y^+}$ (positive Yamabe class), with $\sigma^{cd} \sigma_{cd} \equiv 0$ and with $\tau \neq 0$ does not map to a solution, while for any conformal data set with $\Sigma^3$ closed, with $\lambda \in \mathcal{Y^+}$, with $\sigma^{cd} \sigma_{cd} \neq 0$ and with $\tau \neq 0$ does map to a solution. The complete results, and consequent parametrization for CMC data, are detailed elsewhere. (See, e.g., \cite{B-I}.)

It is not surprising that achieving this goal of a complete parametrization is much more difficult for non CMC solutions. In the CMC case, equations \bref{Lichnero}-\bref{confmom} decouple, and (in the vacuum case) solutions of \bref{confmom} are trivial, so the analysis focuses on the single elliptic PDE (the Lichnerowicz equation) 
\begin{equation}
\Delta \phi = \frac{1}{8} R \phi - \frac{1}{8} \sigma^{cd} \sigma_{cd} \phi^{-7} + \frac{1}{12} \tau^2 \phi ^5.
\label{lichnero}
\end{equation}
For non CMC data, however, one has to work with the fully coupled system \bref{Lichnero}-\bref{confmom}.

One of the keys to the majority of the results which have been obtained for near-CMC data, both here and prior to this work, is the pointwise estimate for $|LW|$ which can be derived from equation \bref{confmom}
\begin{equation}
|LW| < C \max_{\Sigma} \phi^6 \max_{\Sigma} |\nabla \tau|
\label{LWEstimate} 
\end{equation}
We show in section 2 how to derive this estimate, whether or not $(\Sigma^3, \lambda_{ab})$ admits a nontrivial Killing field, and very briefly discuss there how Moncrief and the first author \cite{I-M:nearCMC} \cite{I:posYam} have used it and a semi-decoupled sequence argument to show that for any set of conformal data of the sort $(\Sigma^3$ closed, $\lambda \in \mathcal{Y^-}$, any $\sigma$, $\tau$ nowhere zero, small $\frac{\max_\Sigma |\nabla  \tau |}{c \min_\Sigma |\tau|})$ or $(\Sigma^3$ closed, $\lambda \in \mathcal{Y^+}, \sigma$ not identically zero, small $ \max_\Sigma | \nabla \tau|)$, there is a unique solution to the equations \bref{HamConstr}-\bref{MomConstr}, and hence a unique corresponding near-CMC solution $(\Sigma, \gamma, K)$ of the Einstein constraints. Then in section 3 we state and prove our main result, which says that for any conformal data of the sort $(\Sigma^3$ closed, $\lambda$ with $R \geq 0$, $\sigma \equiv 0, \tau =T + \eta$ with $T$ constant and small $\frac{\max_\Sigma |\nabla \eta|}{|T|}$), equations \bref{Lichnero}-\bref{confmom} admit no solution. We discuss in section 4 a curious result of Alan Rendall, in which he shows that for a certain special class of non CMC conformal data, either no solution exists to \bref{Lichnero}-\bref{confmom}, or the solutions are not unique. We make some concluding remarks in section 5.

\section{ $|LW|$ Estimates and Some Existence Results}

To obtain the desired pointwise estimates for the quantity $|LW|$, we start by considering the model equation 
\begin{equation} 
\nabla^a(LX)_{ab} = J_a
\label{DivL}
\end{equation} 
on a smooth closed Riemannian manifold $(\Sigma^3, \lambda_{ab})$ with a specified one-form field $J_b$. The operator on the left-hand side of this equation, $\nabla L$, is elliptic and self-adjoint. If in addition $(\Sigma^3, \lambda_{ab})$ admits no nontrivial conformal Killing field, then $\nabla L$ is invertible. It then follows from standard elliptic PDE theory \cite{Be} that if $J_b$ is contained in the Sobolev space  $H^p_k(\Sigma^3)$ with $p \geq 2$, then there exists a unique solution $X$ to \bref{DivL}, with $X \in H^p_{k+2}(\Sigma^3)$. Further, the solution $X$ satisfies the inequality
\begin{equation} 
\|X\|_{H^p_{k+2}} \leq c \|J\|_{H^p_{k}}
\label{EllipEst}
\end{equation}
where $c$ is a constant\footnote{Such constants appear in many of the inequalities discussed in this section. Although these constants need not be the same, for convenience we  list them all as ``c"} depending only on $p, k$ and the geometry of $(\Sigma^3, \lambda_{ab})$.

We want to control the absolute value of $|LX|$ using these Sobolev norms on $X$. To do that, we first invoke the Sobolev embedding theorem, which says (in 3 dimensions) that for $p\geq1$ and for $k\geq 0$, if $X \in H^p_k(\Sigma^3)$, then for any integer $l$ and any number $\alpha \in (0,1)$ which satisfy $l + \alpha < k - \frac{3}{p}$, one has $ X \in C^{l, \alpha} ({\Sigma^3)}$. It tells us as well that, for some constant $c$, 
 \begin{equation}
 \|X\|_{C^{l, \alpha}} \leq  c\|X\|_{H^p_k(\Sigma^3)}
 \label{SobEmb}
 \end{equation}
If we assume that $p> 3$ and $J \in H^p_k(\Sigma^3)$, then combining \bref{EllipEst} and \bref{SobEmb} we have that, 
\begin{equation} 
 \|X\|_{C^{k +1, \alpha}} \leq c \|J\|_{H^p_{k}}.
 \label{Estim}
 \end{equation}
 Now, it follows from the definition of the operator $L$ and from the definition of these H\"older norms that
 \begin{equation}
 \|LX\|_{C^0} \leq c \| \nabla X \|_{C^0} \leq c \|X\|_{C^{1, \alpha}}.
 \label{norms}
 \end{equation}
 We also have that
 \begin{equation}
 |LX(x)| \leq   \|LX\|_{C^0}   
 \label{C0}
 \end{equation}
 for all $x \in \Sigma^3$. Thus, setting $k=0$ and combining \bref{Estim}-\bref{C0}, we have
 \begin{equation}
 |LX(x)| \leq c \|J\|_{L^p}
 \end{equation} 
 If we assume that $J$ is continuous, then using the definition of the $L^p(\Sigma^3)$ norm along with the finiteness of the measure on the closed manifold $\Sigma^3$, we obtain finally
 \begin{equation}
  |LX(x)| \leq c \max_{\Sigma^3} |J|.
  \label{PtwiseEstimate}
  \end{equation}
  
Clearly if we prescribe sufficiently smooth conformal data $(\Sigma^3, \lambda_{ab}, \sigma_{cd}, \tau)$ with $(\Sigma^3, \lambda_{ab})$ admitting no conformal Killing fields, and if we make sufficient a priori assumptions  concerning the smoothness of $\phi$, then the argument just described allows us to derive the $|LW|$ estimate \bref{LWEstimate} from \bref{confmom}. What happens if $(\Sigma^3, \lambda_{ab})$ admits a nontrivial conformal Killing field? We now show, working with the model equation \bref{DivL}, that we still obtain this estimate. 

\begin{lemma} Let $\Sigma^3$ be a closed manifold and let $\lambda_{ab}$ be a smooth Riemannian metric on $\Sigma^3$. If the one-form field $J_b$ is continuous on $\Sigma^3$ and satisfies the condition
$\int_{\Sigma^3} V^bJ_b = 0$ for any conformal Killing field $V$ of  $(\Sigma^3, \lambda_{ab})$, then there exists a solution $X$ of equation \bref{DivL}, and every such solution satisfies the estimate  \bref{PtwiseEstimate}.
\label{CKVEstimate}
\end{lemma}
\textit{Proof \footnote{We thank Jack Lee for useful discussions concerning this proof.}}: If $(\Sigma^3, \lambda_{ab})$ admits a nontrivial conformal Killing field, then the operator $\nabla L$ is no longer invertible. However it remains elliptic and self-adjoint. Hence it follows from standard elliptic theory \cite{Be} that equation \bref{DivL} admits a solution $X$ so long as $J_b$ is $L^2$ orthogonal to the kernel of $\nabla L$, and further that any such a solution satisfies the elliptic estimate (for some constant $C$)
\begin{equation} 
\|X\|_{H^p_2} \leq  c ( \|J\|_{H^p_0} + \|X\|_{H^p_0}).
\label{EllipEstimker}
\end{equation} 
One readily verifies that the kernel of the operator $\nabla L$ consists of the conformal Killing fields of  $(\Sigma^3, \lambda_{ab})$, since by definition every conformal Killing field $V$ satisfies $LV=0$, and since if $\nabla L Z=0$, then $ 0 = ( Z, \nabla L Z)_{L^2} = (L Z, L Z)_{L^2}$, which implies (using elliptic regularity) that  $L Z = 0.$ Thus the condition $\int_{\Sigma^3} V^bJ_b = 0$ implies that equation \bref{DivL} admits a solution $X$.

To obtain the desired estimate for $LX$, we need to work with the term $\|X\|_{H^p_0}$ in \bref{EllipEstimker}. We first note that the operator $\nabla L$, viewed as a map from $L^p(\Sigma^3)$ to itself, is Fredholm. It follows from Berger-Ebin splitting \cite{B-E} that there is a closed subspace $Y \subset L^p(\Sigma^3)$ such that $L^p = Ker (\nabla L) \oplus Y = Ker L \oplus Y.$  We now want to argue that for any vector field $V \in Y \cap H^p_2$, we have 
\begin{equation} 
\|V\|_{H^p_0} \leq c \| \nabla LV\|_{H^p_0}.
\label{Lpcontrol}
\end{equation}
 
 Suppose that this is not the case. Then there exists a sequence $\{V_i\} \subset Y\cap H^p_2$ such that $\|V_i\|_{H^p_0} =1$ and $ \|\nabla L V_i\|_{H^p_0} \rightarrow 0.$ It follows from \bref{EllipEstimker} that $\|V_i\|_{H^p_2}$ is bounded; then by the Rellich lemma, one knows that there is a subsequence of $\{V_i\}$ which converges in $L^p$. Let us call this limit $V_\infty$. Since the space $Y$ is closed, $V_\infty \in Y$. By continuity, $\|V_\infty \|_{H^p_0} =1$. Now let $Z$ be any smooth vector field. We calculate\footnote{The pairings here can be interpreted distributionally, as duality pairings between $L^p$ and $L^{p*}$, or as ordinary integrals.} 
 \begin{equation}
 (V_\infty, \nabla L Z) = \lim_{i \rightarrow \infty} (V_i, \nabla L Z) = \lim_{i \rightarrow \infty}(\nabla L V_i, Z)= 0,
 \label{stringeqn}
 \end{equation}
where the last equality is a consequence of the assumption $ \|\nabla L V_i\|_{H^p_0} \rightarrow 0.$ This equation tells us that $V_\infty$ is a weak solution to the equation $\nabla L V_\infty=0.$ It immediately follows from elliptic regularity that $V_\infty$ is contained in the kernel of $\nabla L$ and therefore in the kernel of $L$. But since the norm of $V_\infty$ is nonzero and since by assumption $V_\infty \in Y$, we have a contradiction. This proves that the estimate \bref{Lpcontrol} holds. 
 
 Now let us consider a solution $X$ of the equation \bref{DivL}.  As a consequence of the splitting $L^p  = Ker L \oplus Y,$ we write $X = X_Y + X_0$, with $X_0 \in Ker L$ and $X_Y \in Y$. Note that elliptic regularity guarantees that $X, X_0$, and therefore $X_Y$ are all contained in $H^p_2$. We calculate 
 \begin{eqnarray}
 \|LX\|_{H^p_1} &=& \|LX_Y\|_{H^p_1} \nonumber \\ 
                             &\leq& c \|X_Y\|_{H^p_2}  \nonumber \\
                             &\leq& c (\|\nabla L X_Y\|_{H^p_0} + \|X_Y\|_{H^p_0} )  \nonumber \\
                             &\leq& c \|\nabla L X_Y\|_{H^p_0} \nonumber \\
                             &=& c \|\nabla L X\|_{H^p_0} \nonumber \\
                             &=& c \|J\|_{H^p_0}.
\label{stringeqn2}
\end{eqnarray}
Combining this inequality with the Sobolev embedding result noted above, and using the hypothesized continuity of $J$ on the closed manifold $\Sigma^3$, we obtain the pointwise estimate \bref{LWEstimate} as desired, even if $(\Sigma^3, \lambda_{ab})$ admits nontrivial Killing fields. $\bigtriangleup$ 

This pointwise estimate for $|LX|$ (for $X$ satisfying \bref{DivL})  is useful in proving that equations \bref{Lichnero}-\bref{confmom} admit solutions for various classes of near-CMC conformal data, since it allows one to control the $LW$ terms in \bref{Lichnero} via inequalities on the conformal data, and consequently apply sub-super solution techniques to \bref{Lichnero}. Our proofs of these existence results \cite{I-M:nearCMC} \cite{I:posYam} rely on working with a sequence of semi-decoupled equations of the form 
\begin{eqnarray}
\Delta\phi_n &=& \frac{1}{ 8}R\phi_n - \frac{1}{ 8}(\sigma +
LW_n)(\sigma + LW_n)\phi_n^{-7}  + \frac{1}{ 12}\tau^2
\phi_n^5,
\label{Lichnero_n}
\\
\nabla (LW_n) &=& \frac{2}{3}\phi_{n-1}^6\nabla \tau, 
\label{confmom_n}
\end{eqnarray}
to be solved for the sequence of fields $(\phi_n, W_n)$. The idea is to show that, starting with a more or less arbitrary choice of  $\phi_0$, one can solve \bref{confmom_n} with $n=1$ for $W_1$, solve \bref{Lichnero_n} with $n=1$ for $\phi_1$, and indeed solve the sequence of equations \bref{confmom_n}-\bref{Lichnero_n} for the sequence $(\phi_n, W_n)$. One then shows that the sequence $(\phi_n, W_n)$ converges, and that this limit is a solution of the coupled system \bref{Lichnero}-\bref{confmom}.

One doesn't need these $|LX|$-type estimates to show that the sequence of solutions $(\phi_n, W_n)$ to \bref{Lichnero_n}-\bref{confmom_n} exists. They are crucial for arguing convergence of the sequence, however. In particular, deriving from \bref{confmom_n} the estimate $|LW_n| < c \max_{\Sigma} \phi_{n-1}^6 \max_{\Sigma} |\nabla \tau|$, and then substituting this inequality into \bref{Lichnero_n} and replacing $n-1$ by $n$, one can argue that for sufficiently small $|\nabla \tau|$ there are upper and lower (positive) bounds for all $\phi_n$, independent of $n$ \cite{I-M:nearCMC}. These uniform bounds, combined with a contraction mapping argument,  lead to the proof of  convergence. 

We note that in our first work on the existence of solutions for near-CMC conformal data \cite{I-M:nearCMC}, we assume that the conformal metric is in the negative Yamabe class. For such metrics on closed manifolds, there are no nontrivial conformal Killing fields; thus the simpler argument for the pointwise estimate for $|LW|$ (appearing in \cite{I-M:nearCMC}, and outlined here in the discussion prior to Lemma 1) is sufficient. For our later results, in which we consider positive Yamabe and zero Yamabe conformal metrics, we need the argument of Lemma 2. 

\section{ Non Existence Results}

Our main result here is the following theorem:

\begin{theorem} Let  ($\Sigma^3, \lambda_{ab}, \sigma_{cd}, \tau)$ be a set of conformal data with $\Sigma^3$ closed,  with the scalar curvature of $\lambda$ non negative, with $\sigma^2$ zero everywhere, and with $\tau= T + \rho$ with T a nonzero constant.  For $\frac{|\nabla \rho|}{|T|}$ sufficiently small, equations \bref{Lichnero}-\bref{confmom} admit no solution.
\end{theorem}
\textit{Proof}
Let us assume that there is a solution $(\phi, W)$ to equations \bref{Lichnero}-\bref{confmom} for conformal data satisfying the hypothesis stated in the theorem. If we substitute the conditions  $\sigma=0$ and $\tau=T+\rho$ into \bref{Lichnero}-\bref{confmom}, then $(\phi, W)$ must satisfy 
\begin{eqnarray}
\Delta\phi &=& \frac{1}{ 8}R\phi - \frac{1}{ 8} (LW^{ab} LW_{ab}) \phi^{-7}  +
 \frac{1}{ 12}(T+ \rho)^2 \phi^5,
\label{Lichnerop}
\\
\nabla_a (LW)^a_b &=& \frac{2}{3}\phi^6\nabla_b\rho 
\label{confmomp}
\end{eqnarray}
If we then use \bref{confmomp} to derive the pointwise estimate $|LW| < c \max_{\Sigma} \phi^6 \max_{\Sigma} |\nabla \rho|$, and substitute that into \bref{Lichnerop},  we have
\begin{equation}
\Delta \phi \geq \frac{1}{8} R \phi -c(\max_{\Sigma} |\nabla \rho|)^2 (\max _{\Sigma} \phi)^{12} \phi^{-7} +\frac{1}{12}  (T+\rho)^2 \phi^5.
\label{LichneroIneq1}
\end{equation}
Since by assumption the scalar curvature is non negative, it follows that
\begin{equation}
\Delta \phi \geq  -c(\max_{\Sigma} |\nabla \rho|)^2 (\max _{\Sigma} \phi)^{12} \phi^{-7} +\frac{1}{12}  (T+\rho)^2 \phi^5.
\label{LichneroIneq2}
\end{equation}

Now consider a point $x_m \in \Sigma$ at which $\phi$ achieves its global maximum. Evaluating \bref{LichneroIneq2} at $x_m$, we obtain
\begin{equation} 
\Delta \phi (x_m) \geq (\max_\Sigma \phi)^5 [ \frac{1}{12} (T+ \rho (x_m))^2 -c \max_\Sigma |\nabla \rho|^2].
\label{LichneroIneq3}
\end{equation}
For $\max_\Sigma |\nabla \rho|$ sufficiently small relative to $|T|$, the right hand side of \bref{LichneroIneq3} is positive. But $\Delta \phi(p_m)$ must be non positive, since $p_m$ is a global (and therefore local) maximum. Hence we have a contradiction, from which it follows that for data satisfying the conditions listed in the hypotheses, there is no solution to \bref{Lichnero}-\bref{confmom}. $\bigtriangleup$

Does this same sort of nonexistence result hold if, instead of requiring that $\lambda$ have scalar curvature $R \geq 0$, we impose the less restrictive condition that  $\lambda$ be contained in the positive or zero Yamabe class? Since the conformal method is not conformally covariant unless the conformal data is CMC \cite{B-I}, such a result is not an automatic consequence of Theorem 1. Indeed, to date it is not clear whether or not it is true. 

On the other hand, one does have a stronger result of this sort if one works with the conformal thin sandwich  approach (CTSA) \cite{Y} rather than the conformal method. The procedure for constructing solutions via the CTSA is very similar to that outlined above for the conformal method: One starts by choosing a set of CTSA data $(\Sigma^3, \lambda_{ab}, U_{cd}, \tau, \eta)$ where $\lambda_{ab}$ is a Riemannian metric, $U_{cd}$ is a trace-free $(\lambda^{ab}\sigma_{ab})$ tensor field, $\tau$ is a function, and $\eta$ is a function. One then seeks to solve 
\begin{eqnarray}
\Delta\phi &=& \frac{1}{ 8}R\phi - \frac{1}{ 8}(A^{ab}
A_{ab})\phi^{-7}  + \frac{1}{ 12}\tau^2 \phi^5,
\label{CTSALichnero}
\\
\nabla_a[(2\eta)^{-1} (LX)^a_b] &=& \nabla_a [(2\eta)^{-1}
U^a_b]  +  \frac{2}{3}\phi^6\nabla_b\tau 
\label{CTSAconfmom}
\end{eqnarray}
for $\phi$ and the vector field $X^a$. (Here $A_{ab} \equiv (2 \eta)^{-1}((LX)_{ab}-U_{ab}$.) 
Finally one combines the CTSA data and the solution $(\phi,X )$ to obtain a set of initial data 
\begin{eqnarray}
\gamma_{ab} &=& \phi^4\lambda_{ab}
\label{CTSAgamma}
\\
K_{ab} &=& \phi^{-2}(2\eta)^{-1}(LX_{ab}-U_{ab}) +
\frac{1}{3}\phi^{4}\lambda_{ab}\tau. 
\label{CTSAK}
\end{eqnarray}which satisfies the constraint equations \bref{HamConstr}-\bref{MomConstr} along with a choice of the lapse function $N=\phi^6 \eta$ and the shift vector $M^a=X^a$ which are used to generate evolution. 

One of the key features of the CTSA is that, unlike the conformal method, it is conformally covariant in the sense that a solution exists for the CTSA data $(\Sigma^3, \lambda_{ab}, U_{cd}, \tau, \eta)$ if and only if it also exists for the conformally related CTSA data $(\Sigma^3, \psi^4 \lambda_{ab}, \psi^{-2}U_{cd}, \tau, \psi^6\eta)$, where $\psi$ is any positive definite function;  and furthermore, the initial data $(\gamma, K)$ and the lapse and shift corresponding to each of these sets of data is identical. Combining this fact with an argument very similar to that used to prove Theorem 2, we can show the following

\begin{theorem} Let $(\Sigma^3, \lambda_{ab}, U_{cd}, \tau, \rho)$ be a set of CTSA data with $\lambda \in \mathcal{Y}^+ \cup \mathcal{ Y}^0$, with $U^2$ zero everywhere, and with $\tau= T + \rho$ with T a nonzero constant.  For $\frac{|\nabla \rho|}{|T|}$ sufficiently small, equations \bref{CTSALichnero}-\bref{CTSAconfmom} admit no solution.
\end{theorem}

\section{ Rendall's Conformal Data for which there is Not a Unique Solution}
 
While the results which we have proven in section 3 provide the first examples of sets of non CMC conformal data and sets of non CMC CTSA data which we know do not map to solutions of the constraint equations, earlier unpublished work of Alan Rendall describes a very special set of conformal data for which either a solution does not exist, or if it exists it is not unique.\footnote{Rendall's study of this set of data was partially motivated by Bartnik's work \cite{Bart} on sets of data (similar to this one) which evolve into spacetimes which contain no constant mean curvature Cauchy surfaces.} The proof does not  determine which of these possibilities holds. To stimulate further understanding of this issue, we describe Rendall's results here. 

\begin{theorem} Let  $(\Sigma^3, \lambda_{ab}, \sigma_{cd}, \tau)$ be a set of conformal data with $\Sigma=S^2 \times S^1$,  with $\lambda=$ (round sphere metric) $\times$ (circle metric), with $\sigma^2 \equiv 0$, and with $\tau= f (x)$, where $x$ is the coordinate on the $S^1$ factor, and $f(-x)=-f(x)$. For such data, equations \bref{Lichnero}-\bref{confmom} either admit no solution, or admit more than one solution. 
\end{theorem}
\textit{Proof} Let us define a pair of groups, $\mathcal{S}$ and $\mathcal{Z}^2$, which act on $\Sigma^3$: $\mathcal{S}$ is the rotation group $SO(3)$ which acts on the $S^2$ component of $\Sigma^3$ in the usual way, leaving $S^1$ invariant, while $\mathcal{Z}^2$ is the reflection group, with the element $\Psi$ reflecting $S^1$ across some central point $p_0 \in S^1$, and leaving $S^2$ invariant. We note that the data $(\lambda_{ab}, \sigma_{cd}+ \lambda_{cd} \tau)$ are invariant under the action of $\mathcal{S}$, the metric $\lambda_{cd}$ is invariant under the action of $\Psi$, and the quantity $ \sigma_{cd}+ \lambda_{cd} \tau$ changes its sign under the $\Psi$ action. It  follows that if there exists a unique solution $(\phi, W^a)$ to \bref{Lichnero}-\bref{confmom} 
for this conformal data, then the reconstituted data $(\gamma_{ab}, K_{cd})$ (See equations \bref{recongamma}-\bref{reconK}.) must be invariant under the $\mathcal{S}$ action, have $\gamma$ invariant under the $\Psi$ action, and have $K$ change its sign under the $\Psi$ action.

So let us assume that indeed there exists a unique solution to \bref{Lichnero}-\bref{confmom}, and let us label the resulting initial data set $(\gamma, K)$. We show now that this leads to a contradiction.  Following Rendall \cite{R}, based on the data $(\gamma, K)$ and on its local spacetime development $g$, we define a pair of quantities (i) $\mathcal{R}(x,t)$, which is equal to the $S^2$ radius at the point $(x,t)$, where $x$ is the coordinate for $S^1$ with $x=0$ at the point $p_0$; and (ii) $m(x,t)=\frac{1}{2} \mathcal{R}(x,t)(1-g(\nabla \mathcal{R}(x,t), \nabla \mathcal{R}(x,t))$. As a consequence of the vacuum Einstein equations, $m$ and $\mathcal{R}$ must  satisfy the equations
\begin{eqnarray}
\label{Req}
\nabla_{\alpha} \nabla_{\beta}\mathcal{R}&=&\frac{m}{\mathcal{R}^2} g_{\alpha \beta} \\
\nabla_{\alpha} m &=&0, 
\label{m0}
\end{eqnarray}
where the indices $\alpha, \beta$ take on the two values $x$ and $t$. It of course follows from \bref{m0} that $m(x,t)$ is a constant; we label this constant $\hat{m}$.

We need to verify the following claim: At the point $(x,t)=(0,0)$, $\nabla \mathcal{R}(0,0)=0$, and ${\mathcal{R}(0,0)}=2 \hat{m}$. The first of these claimed equations follows from the facts that (i) since $K_{cd}$ changes sign under the action of $\Psi$, we have $K_{cd}(0,0) =0$ and therefore $\partial_t \mathcal{R}=0$; and (ii) since the metric is invariant under the reflection $\Psi$, $\partial_x \mathcal{R}(0,0)=0$. Thus we have $\nabla_{\alpha} \mathcal{R}(0,0)=0$. The second equation follows immediately from the first, along with the definition of $m$: We have $\hat{m}=m(0,0)=\frac{1}{2} \mathcal{R}(0,0)(1-g(\nabla \mathcal{R}(0,0), \nabla \mathcal{R}(0,0))= \frac{1}{2} \mathcal{R}(0,0)$.

We next consider a global maximum point $x_m \in S^1$ of the function $\mathcal{R}(x,0)$. Since the data is presumed to be smooth, we have $\partial_x\mathcal{R} (x_m,0)=0$. Thus we find that 
\begin{equation}
g(\nabla \mathcal{R}(x_m,0), \nabla \mathcal{R}(x_m,0))= g^{tt}(\partial_t \mathcal{R}(x_m,0))^2 \leq 0. 
\end{equation}
So we have, from the definition of $m$, 
\begin{equation}
\hat{m}=m(x_m,o)=\frac{1}{2} \mathcal{R}(x_m,0)(1-g(\nabla \mathcal{R}(x_m,0),\nabla \mathcal{R}(x_m,0)) \geq \frac{1}{2}\mathcal{R}(x_m,0).
\end{equation}
Since $x_m$ is a global maximum for $\mathcal{R}(x,0)$, it follows from this result that for all $x\in S_1$, 
\begin{equation} 
\mathcal{R}(x,0)\leq R(x_m,0) \leq 2\hat{m}.
\label{xmineq}
\end{equation}

Now, comparing the result (\ref{xmineq}) with our earlier determination that ${\mathcal{R}(0,0)}=2 \hat{m}$, we verify that $x=0$ is a global maximum for $\mathcal{R} (x,0)$. However, using the fact that $K$ vanishes at $(0,0)$, together with the Einstein equation \bref{Req}, we calculate 
\begin{equation}
\partial_x \partial_x \mathcal{R}(0,0)=\nabla_x \nabla_x \mathcal{R}(0,0)=\frac{\hat{m}}{\mathcal{R}^2} g_{xx} (0,0) >0.
\end{equation}
This contradicts our contention that $(0,0)$ is a global maximum for $\mathcal{R}$, completing our proof by contradiction.\footnote{One could also prove this result using the fact that, as a consequence of Birkhoff's theorem, if there were a solution to the constraints based on the choice of conformal data under discussion here, then the spacetime development of this data would necessarily be a portion of the Schwarzschild spacetime (inside the horizon).} $\bigtriangleup$

\section{Conclusion}

In a certain sense, our main result (Theorem 2) is a stability  result for the nonexistence of solutions to  \bref{Lichnero}-\bref{confmom} for certain sets of conformal data. Specifically, we recall that for CMC conformal data of the type $(\Sigma^3$ closed, $\lambda \in \mathcal{Y}^+ \bigcup \mathcal{Y}^0, \sigma^2$ identically zero, $\tau \neq 0)$, there exist no solutions. \cite{I:cmc}. Restricting this result to those special cases in which $R(\lambda)\geq0$, we see that our new results shows that if we perturb the conformal data above by allowing $\tau$ to be non constant with small gradient, then the nonexistence condition still holds. We do not expect to retain nonexistence if we also perturb $\sigma^2$ away from zero, since we know that CMC conformal data of the type $(\Sigma^3$ closed, $\lambda \in \mathcal{Y}^+ \bigcup \mathcal{Y}^0, \sigma^2$ not identically zero, $\tau \neq 0)$ does lead to the existence of unique solutions.

We might wish to see if other sets of CMC conformal data for which no solutions exist are stable in a similar sense. In all other such classes of CMC conformal data, the function $\tau$ is zero. (See the table in section 2 of \cite{I:cmc}.)  The analysis of non CMC conformal data with $\tau$ having zeroes and also having a small gradient has thus far proven difficult. So this problem is still open. 

Also wide open is the question of whether solutions exist for conformal data with no restriction on the gradient of $\tau$. The only known result concerning such data is that of Rendall, described above. New ideas are likely needed to make progress in addressing and answering this question.

\section{Acknowledgments}

We thank the Caltech Numerical Relativity Visitors Program, which hosted us while this research was being done. JI is partially supported by  NSF grant PHY 0099373 at the University of Oregon.


\begin{thebibliography}{10}

\bibitem{CB-Y}
Y.~Choquet-Bruhat and J.~York.
\newblock The {C}auchy problem.
\newblock In A.~Held, editor, {\em General Relativity and Gravitation}. Plenum,
  1980.

\bibitem{I:cmc}
J.~Isenberg.
\newblock Constant mean curvature solutions of the {E}instein constraint
  equations on closed manifolds.
\newblock {\em Class. Quant. Grav.}, 12:2249, 1995.

\bibitem{A-C}
L.~Andersson and P.T. Chru\'sciel.
\newblock On asymptotic behavior of solutions of the constraint equations in
  general relativity with ``hyperboloidal boundary conditions''.
\newblock {\em Dissert. Math.}, 355:1--100, 1996.

\bibitem{B-C}
D.~Brill and M.~Cantor.
\newblock {\em Composit. Math.}, 43:317, 1981.

\bibitem{I-M:nearCMC}
J.~Isenberg and V.~Moncrief.
\newblock A set of nonconstant mean curvature solutions of the {E}instein
  constraint equations on closed manifolds.
\newblock {\em Class. Quant. Grav.}, 13:1819, 1996.

\bibitem{I-P}
J.~Isenberg and J.~Park.
\newblock Asymptotically hyperbolic non-constant mean curvature solutions of
  the Einstein constraint equations.
\newblock {\em Class. Quant. Grav.}, 14:A189, 1997.

\bibitem{CB-I-Y}
Y.~Choquet-Bruhat, J.~Isenberg, and J.~York.
\newblock Einstein constraints on asymptotically euclidean manifolds.
\newblock {\em Phys. Rev. D}, 61:084034, 2000.

\bibitem{B-I}
R.~Bartnik and J.~Isenberg.
\newblock The Constraint Equations
\newblock In P.~Chrusciel and H.~Friedrich, editors {\em 50 Years of the Cauchy Problem}, 2004.

\bibitem{B-E}
M.~Berger and D.~Ebin
\newblock Some Decompositions of the Space of Symmetric Tensors on a Riemannian Manifold.
\newblock {\em J. Diff. Geom. }, 3:379-392, 1969.

\bibitem{I:posYam} 
J.~Isenberg
\newblock unpublished.

\bibitem{Be}
A.~Besse.
\newblock {\em Einstein manifolds}.
\newblock Springer, 1987.

\bibitem{Y}
J.~W. York.
\newblock Conformal {T}hin-{S}andwich data for the initial-value problem of
  general relativity.
\newblock {\em Phys. Rev. Lett.}, 82:1350--1353, 1999.

\bibitem{Bart}
R.~Bartnik
\newblock Remarks on cosmological spacetimes and constant mean curvature surfaces.
\newblock{\em Comm. Math. Phys.}, 117: 615-624, 1988. 

\bibitem{R}
A.~Rendall
\newblock Crushing singularities in spacetimes with spherical, plane, and hyperbolic symmetry.
\newblock {\em Class. Qtm. Grav}, 12:1517, 1995. 
\end{thebibliography}
 \end{document}